\documentclass[prl,aps,twocolumn,nopacs,superscriptaddress,nofootinbib]{revtex4-2}

\usepackage{amsmath}
\usepackage{amssymb}
\usepackage{amsfonts}
\usepackage{bm}
\usepackage{bbm}
\usepackage{braket}
\usepackage{color}
\usepackage{comment}
\usepackage{dcolumn} 
\usepackage{dsfont}
\usepackage{enumerate}
\usepackage{epsfig}
\usepackage{esint}
\usepackage[T1]{fontenc}
\usepackage{framed}
\usepackage{gensymb}
\usepackage{graphicx} 
\usepackage[colorlinks,linkcolor=blue,citecolor=blue,urlcolor=blue,hyperindex,driverfallback=dvipdfm]{hyperref}
\usepackage{indentfirst}
\usepackage{lmodern}
\usepackage{mathrsfs}
\usepackage{mathtools}
\usepackage{multirow}
\usepackage{psfrag}
\usepackage{pst-all}
\usepackage{soul}
\usepackage{xcolor}
\usepackage{xspace}
\usepackage[normalem]{ulem}
\usepackage{soul,xcolor}
\usepackage{orcidlink}

\newcommand{\abs}[1] {\mathopen{}\left|#1\right|\mathclose{}}

\newcommand{\sqpar}[1] {\mathopen{}\left[#1\right]\mathclose{}}
\newcommand{\clpar}[1] {\mathopen{}\left\{#1\right\}\mathclose{}}

\def\ii{{\rm i}}  \def\ee{{\rm e}}
  \def\Imm{{\rm Im}}

\def\rb{{\bf r}}  \def\Rb{{\bf R}}
    
    \def\zz{\hat{\bf z}}

\def\kb{{\bf k}}  \def\kpar{k_\parallel}
\def\Qb{{\bf Q}}
\def\jb{{\bf j}}
    
\def\jb{{\bf j}}    \def\pb{{\bf p}}
\def\rp{r_{\rm p}}    
\def\vF{v_{\rm F}}      
    
\def\eps{\epsilon}  \def\vep{\varepsilon}  \def\ww{\omega}

  \def\Gm{\mathcal{G}}    
  
\def\dperp{{d_\perp}}  \def\dpar{{d_\parallel}}  
\def\dperpSRM{d_{\perp}^{\text{SRM}}}

\def\epsd{\epsilon_{\rm d}}  \def\epsm{\epsilon_{\rm m}}
\def\epsb{\epsilon_{\rm b}}  \def\epsL{\epsilon_{\rm L}}
\def\kpar{{k_\parallel}}
\def\kd{k_{\rm d}}  \def\kdz{k_{{\rm d},z}}
  \def\kmz{k_{{\rm m},z}}

\begin{document}

\title{Nonlocal effects in atom-plasmon interactions}

\author{Mikkel~Have~Eriksen\,\orcidlink{0000-0002-0159-0896}}
\affiliation{POLIMA---Center for Polariton-driven Light--Matter Interactions, University of Southern Denmark, Campusvej 55, DK-5230 Odense M, Denmark}

\author{Christos Tserkezis\,\orcidlink{0000-0002-2075-9036}}
\affiliation{POLIMA---Center for Polariton-driven Light--Matter Interactions, University of Southern Denmark, Campusvej 55, DK-5230 Odense M, Denmark}

\author{N.~Asger~Mortensen\,\orcidlink{0000-0001-7936-6264}}
\affiliation{POLIMA---Center for Polariton-driven Light--Matter Interactions, University of Southern Denmark, Campusvej 55, DK-5230 Odense M, Denmark}
\affiliation{Danish Institute for Advanced Study, University of Southern Denmark, Campusvej 55, DK-5230 Odense M, Denmark}

\author{Joel~D.~Cox\,\orcidlink{0000-0002-5954-6038}}
\email[Joel~D.~Cox: ]{cox@mci.sdu.dk}
\affiliation{POLIMA---Center for Polariton-driven Light--Matter Interactions, University of Southern Denmark, Campusvej 55, DK-5230 Odense M, Denmark}
\affiliation{Danish Institute for Advanced Study, University of Southern Denmark, Campusvej 55, DK-5230 Odense M, Denmark}

\begin{abstract}
Nonlocal and quantum mechanical phenomena in noble metal nanostructures become increasingly crucial when the relevant length scales in hybrid nanostructures reach the few-nanometer regime. In practice, such mesoscopic effects at metal-dielectric interfaces can be described using exemplary surface-response functions (SRFs) embodied by the Feibelman $d$-parameters. Here we show that SRFs dramatically influence quantum electrodynamic phenomena---such as the Purcell enhancement and Lamb shift---for quantum emitters close to a diverse range of noble metal nanostructures interfacing different homogeneous media. Dielectric environments with higher permittivities are shown to increase the magnitude of SRFs calculated within the specular-reflection model. In parallel, the role of SRFs is enhanced in nanostructures characterized by large surface-to-volume ratios, such as thin planar metallic films or shells of core-shell nanoparticles. By investigating emitter quantum dynamics close to such plasmonic architectures, we show that decreasing the width of the metal region, or increasing the permittivity of the interfacing dielectric, leads to a significant change in the Purcell enhancement, Lamb shift, and visible far-field spontaneous emission spectrum, as an immediate consequence of SRFs. We anticipate that fitting the theoretically modelled spectra to experiments could allow for experimental determination of the $d$-parameters.
\end{abstract}

\date{\today}
\maketitle


\section{Introduction}

Nanoscale light-matter interactions, and the strong coupling of light with atoms and molecules, are abundant sources of fundamental physical insights, while offering applications in fields such as optical sensing, photocatalysis, and quantum optics~\cite{zhang2013plasmonic,streltsov2017colloquium,dombi2020strong}. In these and other areas, metallic nanostructures that support plasmons---the collective oscillations of conduction electrons---are sought as light-focusing elements that enhance the interaction of atomic systems with external optical fields~\cite{mertens_apl89,yan_prl98,necada_ijmpb31,cox2018nonlinear}. Meanwhile, in a quantum-electrodynamical context, plasmonic nanostructures are actively explored as subwavelength optical cavities that control the generation of single photons~\cite{chang2006quantum,Bozhevolnyi:2017a,bozhevolnyi2017case,Fernandez-Dominguez2018}, a key resource for future quantum optical information and communication technologies~\cite{obrien_natphot3}. Steady progress in the aforementioned frontier research topics has relied crucially on the framework of classical electrodynamics within the local-response approximation (LRA)~\cite{Mortensen2021review}. However, as the fabrication of plasmonic nanostructures such as metallic nanoparticles (NPs) becomes increasingly advanced, such that the feature size and distance between structures approach the few-nanometer regime, the LRA can no longer accurately estimate the response of the system, as it neglects nonlocal and quantum mechanical corrections in the bulk and at the surface of the metal~\cite{pitarke2006theory,scholl2012quantum,raza2015nonlocal,echarri2019quantum,Mortensen2021review}.

Nonlocal and quantum mechanical phenomena at metal surfaces have been modelled through a series of methods such as descriptions of the bulk response through semi-classical hydrodynamic models~\cite{Raza2011,mortensen2014generalized,toscano2015resonance,ciraci_prb93,trugler_ijmpb31,kupresak_advts3,eremin_jqsrt277,mystilidis_comphy284,wegner2023halevi} or \emph{ab-initio} methods such as time-dependent density-functional theory (TDDFT)~\cite{zuloaga_nl9,teperik_prl110,varas2016quantum,zhu2016quantum}. The standard hydrodynamic Drude model (HDM) that relies on the Thomas--Fermi theory with hard-wall boundary conditions describes the motion of the compressible electron gas as a convective fluid and needs amendments to account for surface phenomena such as electron spill-out or spill-in~\cite{toscano2015resonance,WeiYan2015,ciraci_prb93}, while TDDFT, constituting a more sophisticated method for modelling quantum mechanical phenomena in plasmonic nanostructures, demands huge computational costs and is practically restricted to few-atom structures. The mesoscopic regime, bridging the micro- and macroscopic realms, necessitates a description that goes beyond the classical LRA and yet is less computationally demanding than \emph{ab-initio} approaches~\cite{stamatopoulou2022omex}.

Including surface-response functions (SRFs)~\cite{feibelman1982surface,Yan:2015,Christensen:2017,Yang:2019} in the otherwise classical or semiclassical constitutive relations when solving Maxwell's equations allows us to take into account quantum mechanical phenomena when electrons are confined to small structures in the mesoscopic regime, while still maintaining the simple classical or semiclassical bulk response functions~\cite{tserkezis2021towards, karanikolas2021quantum, Dezfouli:2017, babaze2022quantum, mortensen2021surface}. Feibelman $d$-parameters are such SRFs that permit the modelling of surface-enabled Landau damping, nonlocality, and electron spill-out or spill-in effects to leading order~\cite{gonccalves2020plasmon,gonccalves2020plasmonics,babaze_nanoph10}. The Feibelman $d$-parameters, $\dperp$ and $\dpar$, are associated with an interface between two materials, and depend on the properties of these two materials constituting the interface. The $d$-parameters are often computed using atomistic or \emph{ab-initio} methods for metal-vacuum interfaces, an approach that is prohibitively time-intensive to tabulate for arbitrary metal-dielectric interfaces. We here focus on noble metal interfaces, where d-band screening and spill-in effects cannot be captured by jellium models~\cite{varas2016quantum}; we thus resort to finding analytical expressions for the $d$-parameters using the specular-reflection model (SRM) in combination with HDM for the longitudinal component of the dielectric tensor~\cite{ford1984electromagnetic, echarri2021optical}.

When positioned near a plasmonic nanostructure, a quantum emitter (QE), such as a quantum dot or an atom, will exhibit altered emission properties associated with the modified local photonic density of states (LDOS)~\cite{scully1997quantum, novotny2012principles, hohenester2020nano, chang2017constructing}.
In the specific case of a sodium (Na) NP or interface, the quantum mechanical and nonlocal corrections captured by SRFs result in clear changes to the light emission spectrum of the QE~\cite{karanikolas2021quantum}, as well as quantum electrodynamical phenomena in the form of the Lamb shift and Purcell enhancement of the QE~\cite{gonccalves2020plasmon,babaze2022quantum}. 
Here, we investigate the influence of the $d$-parameters in plasmonic nanostructures comprised of noble metals such as gold (Au) or silver (Ag). We show that the influence of the SRFs on the emission properties of a nearby QE increases when the permittivity of the interfacing dielectric increases, or the surface area becomes larger compared to the volume of the nanostructure by---for example---increasing the number of interfaces when going from a single extended interface to a thin film, or analogously by substituting a solid spherical NP with a core-shell NP. Our results showcase configurations and situations where the inclusion of quantum mechanical corrections in the mesoscopic regime is of particular importance. These large non-classical corrections might allow for experimental determination of the $d$-parameters by, e.g., fitting the theory presented here to experimental results~\cite{gonccalves2023interrogating}.

\section{Results and discussion}

\begin{figure*} \includegraphics[width=\linewidth]{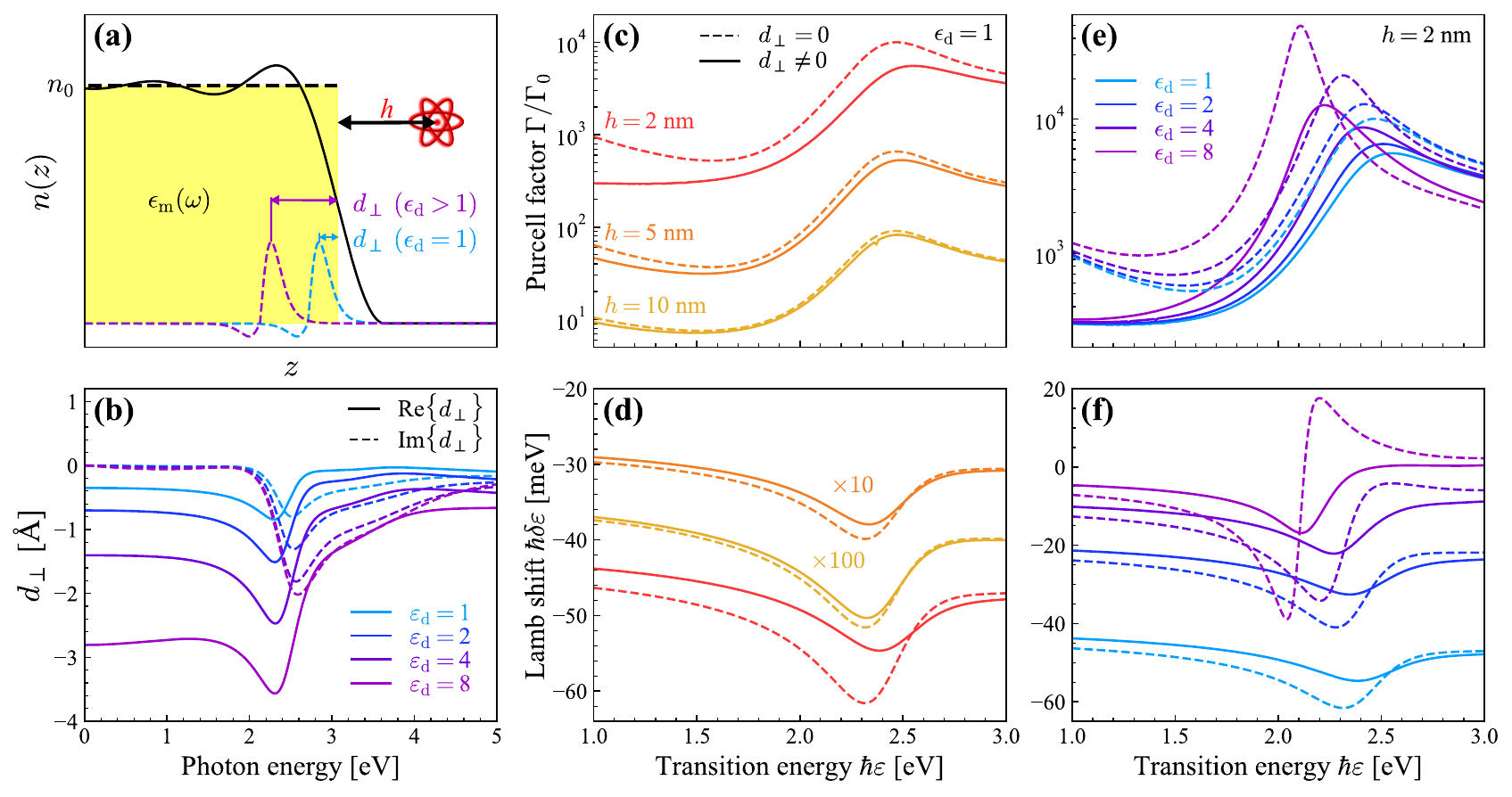}
    \caption{\textbf{The effect of Feibelman $d$-parameters from SRM.} (a) Schematic of a QE (e.g., an atom) located a distance $h$ from a metal with permittivity $\epsm$ interfacing a dielectric material with permittivity $\epsd$. The electron-density profile $n(z)$ (black solid curve) is calculated using a quantum infinite-barrier model~\cite{ford1984electromagnetic}, while the induced-charge density (dashed curves, for the two different $\epsd$ indicated in the legend) and the associated perpendicular $\dperp$-parameters qualitatively show the spill-in depending on $\epsd$. (b) Real (solid curves) and imaginary (dashed curves) parts of the Feibelman $\dperp$-parameter computed from Eq.~\eqref{eq:d_SRM_analytical} for dielectric media with permittivity $\epsd$ (indicated by the color-coded legend) interfacing the metal Au with permittivity $\epsm$ characterized by the model of Ref.~\cite{etchegoin2006analytic}. Panels (c-f) show the Purcell factor $\Gamma/\Gamma_0$ and Lamb shift $\delta\vep$ of the emitter as a function of its transition energy $\hbar\vep$ when omitting (dashed curves) and including (solid curves) the Feibelman $d$-parameters in the metal film response; the effects of separation $h$ (for an emitter in vacuum) and dielectric environment $\epsd$ (for a fixed separation $h=2$\,nm) are explored in panels (c,d) and (e,f), respectively, for an emitter with transition dipole moment $d = 1$\,$e\cdot$nm.}
    \label{fig1}
\end{figure*}

\textbf{Surface response functions for arbitrary metal-dielectric interfaces.} For a particular dielectric--metal interface spanning the $\Rb=(x,y)$ plane, the associated Feibelman $d$-parameters can be obtained regarding the quantum mechanical induced charge and current density in the metal, $\rho(\rb)=\rho(z) \ee^{\ii\Qb\cdot\Rb}$ and $\jb(\rb)=\jb(z)\ee^{\ii\Qb\cdot\Rb}$, respectively, with $Q$ being the in-plane wavevector, from which the parameters are found as~\cite{gonccalves2020plasmon, gonccalves2020plasmonics, feibelman1982surface}:
\begin{subequations}
\begin{align}
    \dperp(\ww) &= \frac{\int_{-\infty}^{\infty} dz z \rho^{\rm ind}(z,\ww)}{\int_{-\infty}^{\infty} dz\rho^{\rm ind}(z,\ww)} , \\
    \dpar(\ww) &= \frac{\int_{-\infty}^{\infty} dz z \partial_z j_\parallel^{\rm ind}(z,\ww)}{\int_{-\infty}^{\infty} dz \partial_z j_\parallel^{\rm ind}(z,\ww)} .
\end{align}
\end{subequations}
In the above expressions, $z$ is taken as the axis normal to the interface, while the system is assumed isotropic in the $xy$ plane. An intuitive physical understanding of the perpendicular parameter, $\dperp$, is here apparent, corresponding to the centroid of the induced charge density as displayed in the schematic of a metal interfacing a dielectric material in Fig.~\ref{fig1}(a), where the equilibrium density of electrons in the metal, $n$, is calculated using a quantum infinite-barrier model and is normalized to the equilibrium density for the infinite electron gas, $n_0$~\cite{ford1984electromagnetic}. The sign of the real part then differentiates between electron spill-out ($\text{Re}\{\dperp\}>0$) or the contrary situation of spill-in ($\text{Re}\{\dperp\}<0$) associated with a high work function.
Correspondingly, $\dpar$ is the centroid of the normal derivative of the in-plane current.

Using these general definitions, one can compute the Feibelman $d$-parameters through \emph{ab-initio} methods, but this requires a new computation for any interface between two different materials. Alternatively, the parameters can be computed using SRM, also known as the semi-classical infinite-barrier (SCIB) model, which assumes specular reflection of the conduction electrons at the interface~\cite{david_jpcc115}. This results in the parameters~\cite{ford1984electromagnetic}:
\begin{subequations}
\begin{align} \label{eq:d_SRM_model}
    & d_\perp^{\text{SRM}} = -\frac{2}{\pi} \frac{\epsm \epsd}{\epsm-\epsd} \int_0^{\infty} \frac{dk_{\rm L}}{k_{\rm L}^2} \left[ \frac{1}{\epsL(k_{\rm L},\ww)} - \frac{1}{\epsm}\right] , \\
    & d_\parallel^{\text{SRM}} = 0 ,
\end{align}
\end{subequations}
where $\dpar=0$ owing to the fact that the interface is intrinsically charge-neutral in the model, $\epsm$ is the classical or semi-classical bulk-metal permittivity, $\epsd$ the corresponding permittivity of the dielectric, and $\epsL$ is the longitudinal permittivity of the spatially dispersive metal---here computed using HDM. Finally, $\omega$ is the angular frequency of the incident light.

In what follows, we use for $\epsm$ a Drude model, while the longitudinal permittivity is obtained within the HDM,
\begin{subequations}
\begin{align}
   \epsm(\ww) &= \epsb(\ww) - \frac{\ww_{\rm p}^2}{\ww^2 + \ii \ww \gamma}, \\
    \epsL(k_{\rm L},\ww) &= \epsb(\ww) - \frac{\ww_{\rm p}^2}{\ww^2 + \ii\ww\gamma - \beta^2 k_{\rm L}^2} ,
\end{align}
\end{subequations}
where $\beta\propto\vF$ contains the dependence on the Fermi velocity $\vF$, being the characteristic velocity of conduction electrons in the metal. In the high (optical) frequency limit $\beta^2 = 3\vF^2/5$~\cite{halevi_prb51,wegner2023halevi}. Finally, $\epsb(\ww)$ is the background permittivity that includes all more complicated contributions such as interband transitions, and is obtained by fitting to experimental data~\cite{johnson1972optical,etchegoin2006analytic}.

With these bulk response functions at hand, an analytical expression for the $d$-parameters can be found in the SRM by using Eq.~\eqref{eq:d_SRM_model}~\cite{svendsen2020role}:
\begin{equation} \label{eq:d_SRM_analytical}
    \dperpSRM = \ii \frac{\epsm \epsd}{\epsm - \epsd} \frac{\beta}{\ww_{\rm p} \sqrt{\epsb}} \left( \frac{\epsb}{\epsm}-1 \right)^{3/2} .
\end{equation}
 
The magnitude and behaviour of the SRFs are investigated by computing $\dperpSRM$ analytically for the noble metal Au, while changing the permittivity of the interfacing dielectric material, in Fig.~\ref{fig1}(b).
For all frequencies under consideration here, $\text{Re}\{\dperp\}<0$, indicating electron spill-in for the conduction electrons in gold, whose work function is relatively high. Such spill-in has previously been associated with core electron screening in noble metals~\cite{liebsch1993surface}, and is at odds with what is typically seen in jellium metals with lower work functions (e.g., Na), which exhibit spill-out and therefore a positive perpendicular Feibelman parameter~\cite{gonccalves2020plasmon}. Fig.~\ref{fig1}(b) also reveals that the magnitude of $\dperpSRM$ becomes larger for dielectric media with higher permittivity, resulting in a pronounced increase of the spill-in at lower frequencies. For instance $\dperpSRM$ approaches $\dperpSRM \approx -0.35$\,$\AA$ at lower frequencies when the dielectric permittivity is $\epsd=1$, but goes towards $\dperpSRM \approx -2.8$\,$\AA$ for $\epsd=8$.

\begin{figure*}
    \centering
    \includegraphics[width=\linewidth]{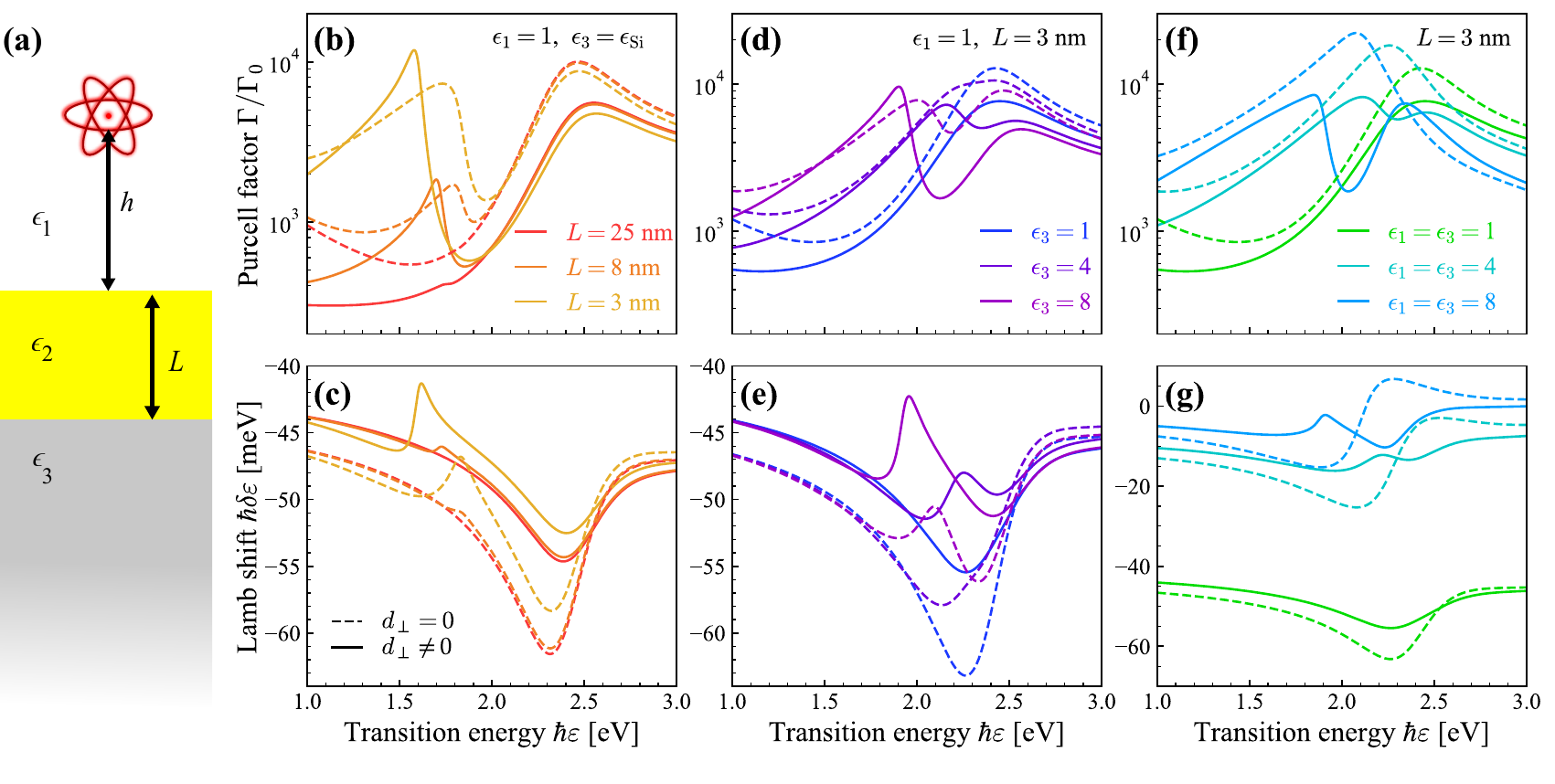}
    \caption{\textbf{Enhanced surface effects in the quantum electrodynamic response of thin metal films.} (a) Schematic of a QE at distance $h$ from a thin metal film of width $L$. In panels (b-g) the calculated Purcell factors and Lamb shifts experienced by a QE with transition dipole moment $d=1$\,$e\cdot$nm placed a distance $h=2$\,nm above a film are shown for cases with (solid curves) and without (dashed curves) incorporating SRFs in the optical response. Panels (b,c) show results for a QE in vacuum ($\eps_1=1)$ close to an Au film of varying thickness, on a Si substrate with permittivity $\eps_3$ interpolated from experimental data~\cite{green2008self}. For a film with $L=3$\,nm, panels (d,e) show the effect of substrate permittivity $\eps_3$ when the QE is in vacuum, while panels (f,g) show the quantum electrodynamic response when the Au film is embedded in a medium with varying permittivity $\eps_1=\eps_3$.}
    \label{fig2}
\end{figure*}

\textbf{Semi-infinite metal films.} The influence of the SRFs can manifest as a change in the emission properties of a QE positioned near a metal film at $\rb=(x,y,h)$, i.e., a distance $h$ from a metallic film extended in the $xy$ plane, as illustrated schematically in Fig.~\ref{fig1}a. Invoking the macroscopic quantum electrodynamics formalism detailed in the SI~\cite{forati2014graphene,ge2013accessing, antao2021two,eriksen2022optoelectronic,scheel2008macroscopic} for a QE characterized as a two-level system with transition energy $\hbar\vep$ and dipole moment $\pb$, the total QE spontaneous emission rate is
\begin{equation} \label{eq:Purcell}
    \Gamma = \Gamma_0 + \frac{2\mu_0}{\hbar}\vep^2\Imm\clpar{\pb^*\cdot\Gm_\vep^{\rm ref}(\rb,\rb)\cdot\pb} ,
\end{equation}
where $\Gamma_0=\vep^3\abs{\pb}^2/3\pi\eps_0\hbar c^3$ is the vacuum decay rate, while the shift in the bare atomic transition frequency due to the photonic environment---the Lamb shift---is
\begin{equation} \label{eq:Lamb}
    \delta \vep = \frac{ \mu_0}{\pi \hbar}\mathcal{P} \int_0^{\infty}d\omega \omega^2  \frac{\Imm\clpar{\pb^*\cdot\Gm_\ww^{\rm ref}(\rb, \rb)\cdot \pb}}{\vep - \omega} ,
\end{equation}
with $\mathcal{P}$ denoting the principal value of the integral. The above expressions depend on the reflected part of the Green's tensor $\Gm_\ww^{\rm ref}$ at the QE location, which quantifies the QE self-interaction mediated by the metal-dielectric interface~\cite{yao2009ultrahigh}. For a dipole oriented perpendicular to the metal surface, i.e., $\pb=p\zz$, the nonvanishing Green's tensor component entering Eqs.~\eqref{eq:Purcell} and \eqref{eq:Lamb} reads as
\begin{equation} \label{eq:G_ref_planar_zz}
    \Gm_{\ww,\perp}^{\rm ref} = \frac{\ii}{4\pi \kd^2}\int_0^\infty d\kpar k_{\parallel}^3\frac{\ee^{\ii 2k_{{\rm d},z}z}}{k_{{\rm d},z}}\rp(\kpar,\ww) ,
\end{equation}
where the Fresnel reflection coefficient for p-polarized light,
\begin{equation} \label{eq:rp_pol_refl_coeff_planar}
    \rp = \frac{\epsm \kdz - \epsd \kmz + (\epsm - \epsd) (\ii k_{\parallel}^2 \dperp - \ii \kdz \kmz \dpar)}{\epsm \kdz + \epsd \kmz - (\epsm - \epsd) (\ii k_{\parallel}^2 \dperp + \ii \kdz \kmz \dpar)},
\end{equation}
depends on the optical wavevector $\kb_j$ in a medium of permittivity $\eps_j$ for $j\in\{{\rm d},{\rm m}\}$, with normal component $k_{j,z}=\sqrt{\eps_j\ww^2/c^2-\kpar^2+\ii0^+}$ and conserved parallel component $\kpar$, as well as on the SRFs that modify the electromagnetic boundary conditions at the metal-dielectric interface.

Using the above expressions, we compute the enhancement in spontaneous emission---quantified by the Purcell factor $\Gamma/\Gamma_0$---along with the Lamb shift $\delta \vep$ for a QE in a medium with permittivity $\eps_1$ that is placed a distance $h$ from a gold interface. In Figs.~\ref{fig1}c and \ref{fig1}d, the Purcell factor and Lamb shift are respectively presented in calculations adopting a fixed vacuum permittivity $\eps_1=1$ at separations $h$ indicated in the legend of Fig.~\ref{fig1}c when omitting (dashed curves) and including (solid curves) SRFs. In the cases considered, the SRFs are found to introduce damping and spectral shifts in prominent plasmonic features of the Purcell factor and Lamb shift that become increasingly important as the QE is brought extremely close to the metal--dielectric interface. The effect of the surface response in quantum electrodynamical phenomena is further amplified when the permittivity of the dielectric medium is increased to enhance spill-in of the interfacing metal electron gas, as revealed in Figs.~\ref{fig1}d and \ref{fig1}e by the change in Purcell factor and Lamb shift, respectively, for a QE with fixed separation $h=2$\,nm and varying environment-permittivity $\eps_1$. The above results underscore the importance of nonclassical surface effects in quantum electrodynamics at a metal-dielectric interface, particularly for dielectric media with high permittivity.

\textbf{Surface response of thin films.} Thin metal films present larger surface-to-volume ratios that should enhance SRF contributions in nanoscale light--matter interactions, while also exhibiting higher sensitivity to screening from interfacing dielectric media. In Fig.~\ref{fig2} we explore the interaction of a QE in a dielectric medium $\eps_1$ placed a distance $h$ above a thin gold film with permittivity $\eps_2$ and thickness $L$ on a substrate with permittivity $\eps_3$, as illustrated schematically in Fig.\ \ref{fig2}a. The film reflection coefficients are
\begin{equation} \label{eq:R_film}
    R_\alpha = r^{(12)}_\alpha + \frac{t^{(21)}_\alpha t^{(12)}_\alpha r^{(23)}_\alpha \ee^{\ii 2 k_{2,z} L}}{1-r^{(21)}_\alpha r^{(23)}_\alpha\ee^{\ii 2 k_{2,z} L}}
\end{equation}
for $\alpha\in\{{\rm s},{\rm p}\}$ polarization, where $r^{(jj')}_\alpha$ and $t^{(jj')}_\alpha$ are reflection and transmission coefficients, respectively, associated with light impinging from a medium with permittivity $\eps_j$ on an interface with permittivity $\eps_{j'}$, which contain the dependence on SRFs as detailed in the SI.

\begin{figure}
    \centering
    \includegraphics[width=\linewidth]{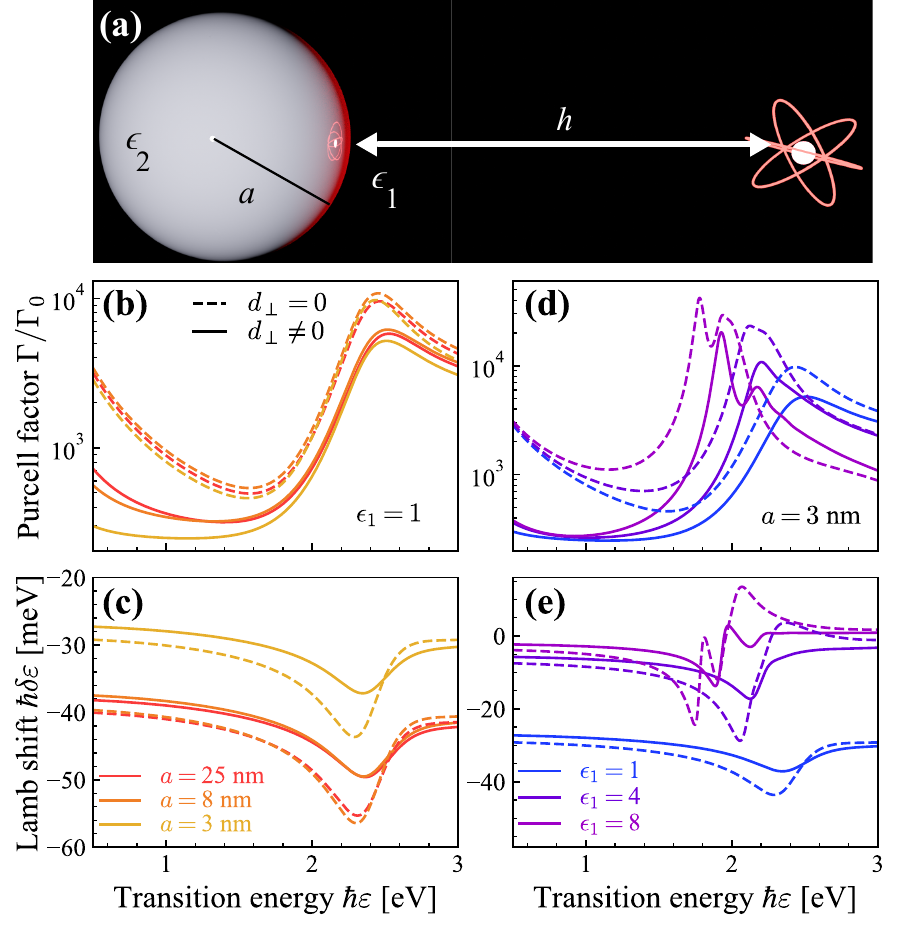}
    \caption{\textbf{Surface effects in quantum light emission near a spherical plasmonic nanoparticle.} (a) Schematic of a QE located a distance $h$ from the surface of a spherical NP with radius $a$. Panels (b,c) show the Purcell factor and Lamb shift of a QE in vacuum ($\eps_{1} = 1$) close to Au NPs with radii indicated in (c), while panels (d,f) show results obtained for a small NP with fixed radius $a=3$\,nm and varying background dielectric constant. Results are obtained for a QE transition dipole moment $d = 1$\,$e\cdot$nm placed at $h=2$\,nm from the outer surface of the NP, with solid and dashed curves indicating results obtained with and without including SRFs, respectively.}
    \label{fig3}
\end{figure}

We explore the influence of SRFs on the Purcell factor and Lamb shift in Figs.~\ref{fig2}b and \ref{fig2}c, respectively, for a QE in vacuum ($\eps_1=1$) above gold films ($\eps_2=\eps_{\rm Au}$) with varying thickness $L$ deposited on silicon ($\eps_3=\eps_{\rm Si}$, obtained from interpolated experimental data in Ref.~\cite{green2008self}) by replacing the single-interface reflection coefficient in Eq.~\eqref{eq:G_ref_planar_zz} with that of Eq.~\eqref{eq:R_film} obtained with (solid curves) and without (dashed curves) SRFs. In thin metal films, the largest deviation from a classical description of the photonic environment is predicted for the thinnest film, where a second plasmonic peak is present at low energies coming from the plasmonic mode at the lower Si-Au interface. 
Interestingly, low-energy features in the Purcell factor and Lamb shift are enhanced by SRFs in the thinnest film, presumably due to the large negative real values of $\dperp$ that describe spill-in of the metal film charge density that effectively reduces the thickness of the free electron gas and boosts plasmonic field confinement. In contrast, high-energy features in the emission spectra experience greater surface damping quantified by the imaginary part of $\dperp$. Similar behavior is seen by considering substrates with different dielectric permittivities, as shown in Figs.~\ref{fig2}d and \ref{fig2}e, where an additional peak emerges when substrates with high permittivities are considered. 
In Figs.~\ref{fig2}(f,g) we consider the impact of SRFs on the QE self-interaction when $\eps_1=\eps_2$ and the metal film is embedded in a high-permittivity dielectric, where plasmon hybridization between the two surfaces may yield a low-energy bonding mode and a higher energy antibonding mode~\cite{Bozhevolnyi:07,raza2013nonlocal}. For $\eps_1=\eps_3=8$ in Fig.~\ref{fig2}f one can see the splitting associated with the hybridization between the two surfaces when including $d$-parameters but not when classically computing the Purcell factor. This can be interpreted as the SRFs effectively decreasing the width of the thin film and hence increasing the hybridization effect through electron spill-in (a negative $\dperp$). This is most pronounced when considering high dielectric environments to the gold film.

\begin{figure*}
    \centering
    \includegraphics[width=\linewidth]{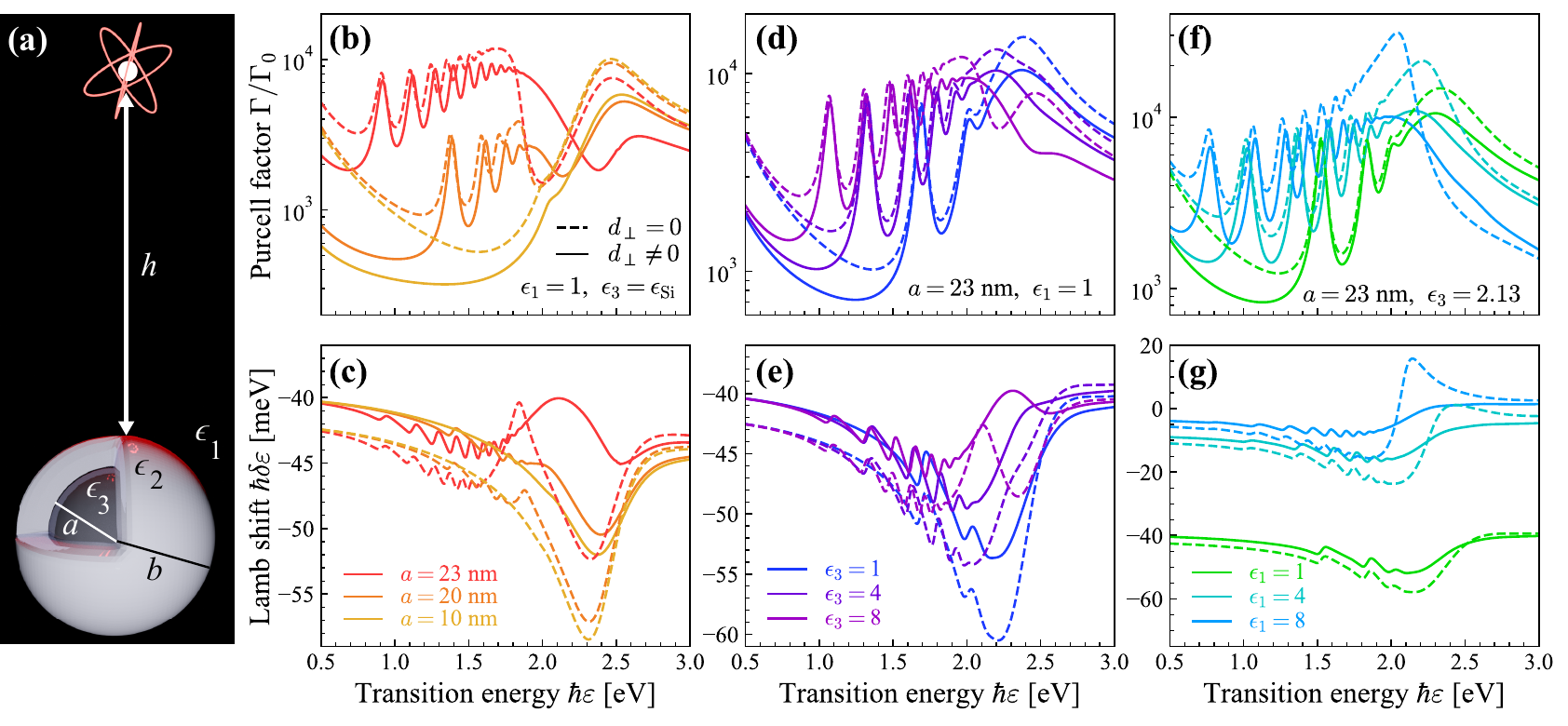}
    \caption{\textbf{Dependence of the Lamb shift and Purcell factor on surface effects in a plasmonic CSNP.} (a) Schematic of a QE at distance $h$ from the surface of a CSNP with core radius $a$ and shell radius $b$. (b,c) show, respectively, the Purcell enhancement and Lamb shift of a QE close to a CSNP with a Si core ($\eps_3$ from Ref.~\cite{green2008self}) and Au shell ($b=25$\,nm) at varying core-radii as indicated in (c). (d,e) consider a core with fixed radius of $a=23$\,nm and varying permittivity. Lastly, in panels (f,g) the core is fixed with $\eps_3=2.13$ to mimic SiO$_2$ and the environment permittivity $\eps_1$ is varied. Results are obtained for a QE with transition dipole moment $d = 1$\,$e\cdot$nm placed a distance $h=2$\,nm from the outer NP surface and the background dielectric material is $\eps_1=1$ except for in (f,g).}
\label{fig4}
\end{figure*}

\textbf{Spherical metallic nanoparticles.} Metallic NPs supporting localized plasmon resonances are quintessential light-focusing elements in nanophotonics that are conveniently described using Mie theory. As depicted in Fig.~\ref{fig3}a, we consider the light emission properties of an emitter characterized by a radially-oriented dipole moment in a medium with permittivity $\eps_1$ at a distance $r$ from the center of a spherical metallic NP with radius $a$ and permittivity $\eps_2$. The quantum electrodynamic response is quantified by the reflected part of the Green's function
\begin{equation} \label{eq:Generalized_Green_fct_zpolarized}
    \Gm_\vep^{\rm ref}(r) = -\frac{\ii k_1}{4\pi}\sum_l l(l+1)(2l+1) a_l \left[\frac{h_l^{(1)}(k_1r)}{k_1r}\right]^2 ,
\end{equation}
where
\begin{widetext}
\begin{gather} \label{eq:Mie_coeff_NP}
    a_l = \frac{\eps_2 \Psi_l^{\prime}(x_1) j_l(x_2) - \eps_1 j_l(x_1) \Psi_l'(x_2)+ (\eps_2-\eps_1) \left[\Bar{d}_\perp j_l(x_1) j_l(x_2) + \Bar{d}_\parallel \Psi_l^{\prime}(x_1)\Psi_l'(x_2)\right]}{\eps_2 j_l(x_2) \xi_l'(x_1)- \eps_1 h_l^{(1)}(x_1) \Psi_l'(x_2)+(\eps_2-\eps_1)\left[\Bar{d}_\perp h_l^{(1)}(x_1)j_l(x_2)+\Bar{d}_\parallel\Psi_l'(x_2)\xi_l'(x_1)\right]}
\end{gather}
\end{widetext}
are Mie coefficients that are linearized in the Feibelman $d$-parameters, as reported in Ref.~\cite{gonccalves2020plasmon}, normalized according to $\Bar{d}_\perp \equiv l(l+1)\dperp/a$ and $\Bar{d}_\parallel \equiv \dpar/a$, while $j_l$ and $h_l^{(1)}$ are spherical Bessel and Hankel-of-the-first-kind functions, respectively, of the normalized parameter $x_j\equiv k_j a$ that also enters the derivatives of the Riccati--Bessel functions $\Psi_l'(x)=\partial_x\sqpar{x j_l(x)}$ and $\xi_l'(x)=\partial_x[x h_l^{(1)}(x)]$.

\begin{figure*}
    \includegraphics[width=\linewidth]{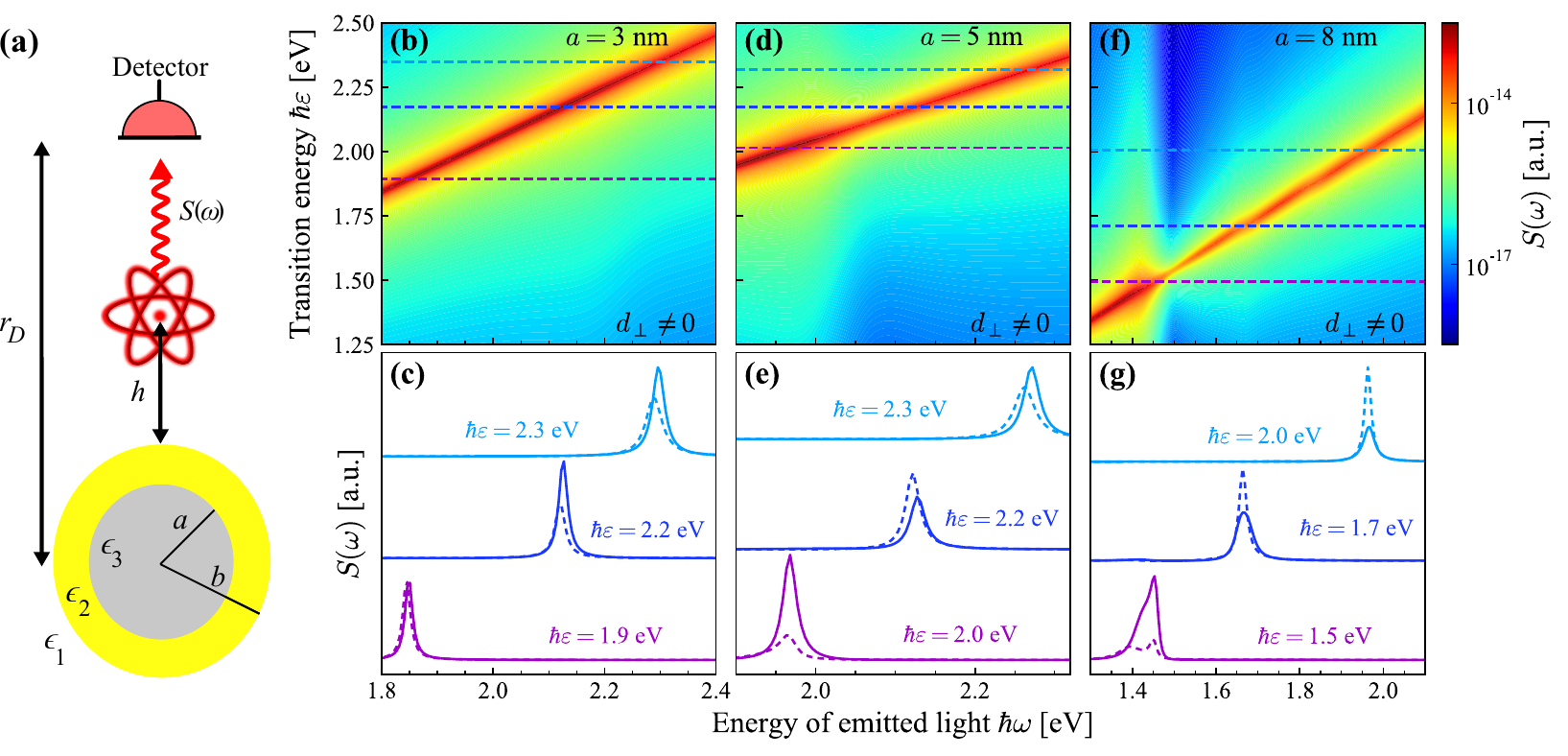}
    \caption{{\bf Detection of surface response effects in spontaneous emission produced near a plasmonic nanostructure.} (a) Schematic illustration of a QE in a medium with permittivity $\eps_1$ at distance $h$ from the outer surface of a CSNP, with shell radius (permittivity) $b$ ($\eps_2$) and core radius (permittivity) $a$ ($\eps_3)$, emitting light that is detected in the far field at a distance $r_{\rm D}$. Panels (b-g) show the spontaneous emission spectra $S(\omega)$ detected at $r_{\rm D}=1$\,$\mu$m for a QE with transition frequency $\vep$, dipole moment $d=1e\cdot$nm, and intrinsic broadening $\hbar\gamma_{\rm d}=15$\,meV (typical for a quantum-dot exciton at room temperature) positioned at $h=2$\,nm from a CSNP with Si core ($\eps_3$ from Ref.~\cite{green2008self}) of radii $a$ indicated in each column and Au shell with radius $b=10$\,nm; the upper row of panels (b,d,f) shows contours of $S(\omega)$ that sweep the detected light energy $\hbar\omega$ on horizontal axes and the QE transition energy on the vertical axes, while the lower row of panels (c,e,g) shows the emission spectrum for specific QE transition energies $\hbar\vep$---indicated by the color-coded dashed horizontal lines in the panels immediately above---when including (solid curves) or omitting (dashed curves) SRFs (each set of curves is appropriately re-scaled for clarity).}
    \label{fig5}
\end{figure*}

With the above expressions at hand, the Purcell enhancement and Lamb shift of a QE near a spherical Au NP are found for a series of radii in Figs.~\ref{fig3}(b,c). The $d$-parameters in Eq.~\eqref{eq:Mie_coeff_NP} go as $l(l+1)\dperp/a$ and $\dpar/a$, meaning that their relative contribution increases for smaller NP radius and higher multipolar modes. As the higher multipolar modes influence the Green's function for smaller distances between the QE and the NP, as seen in Eq.~\eqref{eq:Generalized_Green_fct_zpolarized}, one would expect the influence of SRFs to increase for smaller radii, in agreement with the results in Figs.~\ref{fig3}(b,c). In addition, when varying the background permittivity rather than the radius of the NP, as in Fig.~\ref{fig3}(d,f), one sees that the larger permittivity ensures a larger deviation from the classical results when including the SRF, an effect which is particularly noticeable in the Lamb shift. Incidentally, as we show in the SI, the above results obtained using Mie theory are qualitatively reproduced by incorporating SRFs in the multipolar polarizability obtained in the quasistatic approximation, which is well-justified when the size of the NP and the QE-NP separation are small. 
    
\textbf{Surface response effects in spherical core-shell nanoparticles.} Compared to spherical NPs, the additional metal-dielectric interface in spherical dielectric core--metal shell NPs (CSNPs) leads to a more involved dependence on surface effects in the optical response, as indicated by the lengthy analytical expressions reported in the SI for both the associated Mie coefficients and the quasistatic multipolar polarizability. For a dipole oriented in the radial direction and placed a distance $h$ from the center of a CSNP with core radius $a$ and shell radius $b$, as illustrated in Fig.~\ref{fig4}a, the Purcell factor and Lamb shift corresponding to the solutions from Mie theory are plotted in Figs.~\ref{fig4}(b-g). The sharper spectral features in the classical estimation of the Purcell factor and Lamb shift are slightly damped and blueshifted by SRFs, which can also produce qualitative changes around these multipolar modes. Here, the $d$-parameters appear in the polarizability as $l(l+1)\dperp/R$ for $R\in \{a,b\}$, with the higher-order modes tending to exhibit larger SRF-contributions resulting in larger blueshift and damping. Similarly to the spherical NP, the influence of SRFs is most prominent in higher-order multipolar modes, with the $d$-parameters corresponding to the different interfaces contributing more when the radius to the corresponding interface is small. This can be seen in Figs.~\ref{fig4}(b,c), where the introduction of $d$-parameters to the CSNP with the thinnest shell generally damps and influences the Purcell enhancement and Lamb shift more as compared to CSNPs with thicker shells. We remark that qualitatively similar results are obtained by introducing SRFs in a quasistatic description of the CSNP response, which conveniently leads to analytical expressions for the multipolar polarizability (see SI).

In Figs.~\ref{fig4}(d,e) and Figs.~\ref{fig4}(f,g) we explore the impact of SRFs in the quantum electrodynamic response of the CSNP when the core permittivity and the surrounding permittivity are varied. In the former case, the oscillations in the Purcell enhancement and Lamb shift of the QE associated with the core-shell interface are significantly blue-shifted and damped in the higher-frequency regime when introducing the SRF for high-permittivity cores. In the latter situation, choosing silicon dioxide (SiO$_2$)--also known as silica--as the core while the surrounding permittivity is varied, the high-frequency peak associated with the shell-surrounding interface more pronounced but similar behaviour is observed by the inclusion of the SRF.

The spontaneous emission spectrum of a QE positioned in a photonic environment can be written in terms of the Green's function characterizing the environment:
\begin{gather} \label{eq:Spectrum}
    S(\rb_{\rm D},\ww) = \left| \frac{\mu_0 \ww^2 \pb^* \cdot \Gm_{\ww}(\rb_{\rm D}, \rb) (\ww+\vep)}{\vep^2-\ww^2 -\ii\ww\gamma_{\rm d} -2\mu_0\vep\ww^2\pb^*\cdot \Gm_{\ww}(\rb,\rb)\cdot \pb/\hbar}  \right|^2 ,
\end{gather}
where $\rb_{\rm D}$ denotes the detector position in the far field. The above expression is found in the weak excitation approximation by using the Wiener--Khinchin theorem and accounting for non-Markovian effects~\cite{meystre2007elements,yao2009ultrahigh,van2012spontaneous}. These spectra allow for the investigation of possible strong coupling regimes accessible in these systems, and of the impact of SRFs on experimental observables in these regimes. 

The spectrum of the QE computed from Eq.~\eqref{eq:Spectrum} is displayed in Fig.~\ref{fig5} for a CSNP with an Au shell with fixed radius $b=10$\,nm and a Si core in Figs.~\ref{fig5}(b-g) with varying radii of the core as described in the figure caption. Silicon is used, since it is a dielectric material with a large permittivity in the frequency window of interest, and any effects from SRFs should be more pronounced than for a lower permittivity material such as SiO$_2$. The $d$-parameters introduce a large visible blue-shift in all cases as seen in Fig.~\ref{fig5}, which is as expected for the gold shell as the $d$-parameters introduces spill-in.

The spectra displayed in Fig.~\ref{fig5} exhibit two peaks at some QE transition frequencies, which can be seen for $\hbar\vep=1.5$\,eV in Fig.~\ref{fig5}(g). These peaks are significantly influenced by the SRFs; the amplitude of the spectrum is here enhanced by the SRFs and the two classical peaks coalesces into one peak. It is clear from Fig.~\ref{fig5} that the inclusion of $d$-parameters influences the spectrum dramatically with a generally more pronounced effect for the thinnest shell in Figs.~\ref{fig5}(g). This supports the claim that SRFs are particularly important for thin films or shells, where the surface-to-volume ratio is larger, allowing the SRFs to play a vital role in the optical response.

\section{Conclusions}

Quantum mechanical and nonlocal phenomena at metal-dielectric interfaces become important in the description of light-matter interactions on $\lesssim10$\,nm length scales. These phenomena impact the quantum electrodynamic response of a quantum emitter in close proximity to noble metal nanostructures, as revealed here by the large differences in the Lamb shift and Purcell factor that emerge when comparing the classical response of metallic nanostructures to the nonlocal and quantum mechanical response found from the Feibelman $d$-parameter formalism. In particular, the $d$-parameters obtained from the SRM predict a sizeable increase in $\dperp$ when high-permittivity media interface the noble metal, which is especially important when regarding complicated nanophotonic systems with multiple interfaces and increased surface-to-volume ratios. This finding is showcased for metal films and CSNPs that exhibit pronounced differences in the Lamb shift and Purcell enhancement of a nearby QE when thinner films or shells are considered. The influence of $d$-parameters on the spontaneous emission spectrum of a QE close to a noble metal nanostructure has similarly been investigated, where a large difference in the spectrum can be seen in the case of Si--Au CSNPs. The QE spectrum may be observed in experiments, and constitutes a route for experimentally determining the $d$-parameters by comparing the experimental results with theoretical modelling, similar to the strategy proposed in Ref.~\cite{gonccalves2023interrogating} for determining $d$-parameters using electron-beam spectroscopy. The analytical results presented here facilitate straightforward evaluation of the spectrum, Purcell enhancement, and Lamb shift of QEs close to commonly explored metallic nanostructure morphologies in both quasistatic and retarded regimes.

\section{Acknowledgements}

We thank P. A. D. Gon{\c{c}}alves and C. Wolff for stimulating discussions.
J.~D.~C. is a Sapere Aude research leader supported by Independent Research Fund Denmark (grant no. 0165-00051B).
N.~A.~M. is a VILLUM Investigator supported by VILLUM FONDEN (Grant No.~16498).
The Center for Polariton-driven Light--Matter Interactions (POLIMA) is funded by the Danish National Research Foundation (Project No.~DNRF165).


%

\end{document}